\documentclass[letter,nofootinbib,12pt]{revtex4-2}

\usepackage[colorlinks,citecolor=blue]{hyperref}
\usepackage{amsmath}
\usepackage{mathrsfs}
\usepackage{bbm}
\usepackage{amsfonts}
\usepackage{amssymb}
\usepackage{latexsym}
\usepackage{graphicx}
\usepackage[english]{babel}
\usepackage{multirow}
\usepackage{float}
\usepackage{url}
\usepackage{slashed}
\usepackage[compat=1.1.0]{tikz-feynman}
\usepackage{orcidlink}

\renewcommand{\Delta}{\varDelta} 
\renewcommand{\Gamma}{\varGamma} 
\renewcommand{\Omega}{\varOmega} 
\renewcommand{\Phi}{\varPhi} 
\renewcommand{\Psi}{\varPsi} 
\renewcommand{\Sigma}{\varSigma} 
\renewcommand{\Theta}{\varTheta} 
\renewcommand{\epsilon}{\varepsilon}

\newcommand{\be}{\begin{equation}}
\newcommand{\ee}{\end{equation}}
\newcommand{\ba}{\begin{array}}
\newcommand{\ea}{\end{array}}
\newcommand{\bea}{\begin{eqnarray}}
\newcommand{\eea}{\end{eqnarray}}

\begin{document}

\title{GUT origins of general electroweak multiplets \\ and their oblique parameters}

\author{Liang Chen
\orcidlink{0000-0002-0224-7598} \, }
\email{bqipd@protonmail.com}
\affiliation{School of Fundamental Physics and Mathematical Sciences,
Hangzhou Institute for Advanced Study, UCAS, Hangzhou 310024, China}
\affiliation{University of Chinese Academy of Sciences, 100190 Beijing, China}
\author{Ta-Wei Chan}
\email{chanqq00@gmail.com}
\affiliation{Department of Physics, National Taiwan Normal University,
Taipei 116, Taiwan} 
\author{Thomas W. Kephart \orcidlink{0000-0001-6414-9590} \, }
\email{tom.kephart@gmail.com}
\affiliation{Department of Physics and Astronomy, Vanderbilt
University, Nashville, TN 37235, USA} 
\author{Wai Yee Keung \orcidlink{0000-0001-6761-9594} \, }
\email{keung@uic.edu}
\affiliation{\small Department of Physics, University of Illinois at Chicago, Illinois 60607 USA}
\author{Tzu-Chiang Yuan  \orcidlink{0000-0001-8546-5031} \,}
\email{tcyuan@gate.sinica.edu.tw}
\affiliation{Institute of Physics,
Academia Sinica, Nangang,
Taipei 11529, Taiwan} 


\begin{abstract}
We survey the scalar electroweak multiplets that can originate from a grand unified theory (GUT) like $SU(5)$, $SO(10)$ or $E_6$.
We compute the oblique parameters $S$, $T$ and $U$ for a general scalar electroweak multiplet and then apply the results to the leptoquark, as well as the color sextet and octet cases, allowed by the GUT survey.
Constraints from precision measurement data of the Standard Model $W$ boson mass and $\rho$ parameter to the mass splittings within each of these multiplets are presented. Extension to general vector electroweak multiplets is briefly discussed.
\end{abstract}


\maketitle

\section{Introduction}
\label{Intro}

Leptoquarks (LQs)~\cite{Dorsner:2016wpm} are color-triplet/antitriplet bosons carrying both baryon number and lepton number. They can couple to Standard Model (SM) particles directly, which puts restrictions on their allowed quantum numbers or indirectly with less strictly assigned quantum numbers~\cite{Dorsner:2016wpm,ParticleDataGroup:2020ssz}. 
The Pati-Salam model~\cite{Pati:1974yy} which extends the SM model color gauge group $SU(3)_C$ to $SU(4)$ was the first Grand Unified Theory (GUT) predicting the existence of a leptoquark state. Other GUTs based on $SU(5)$~\cite{Georgi:1974sy}, $SO(10)$~\cite{Georgi:1974my} or $E_6$~\cite{Candelas:1985en,Ma:1987zm,Gunion:1987ge} also made the forecast of scalar or vector LQ states. LQs can also appear in the first stage of symmetry breaking in unified theories based on even larger gauge groups such as $SU(N)\times SU(N)$ ($N$ up to 16)~\cite{Fritzsch:1974nn}.

LQs have been searched for and studied in the context of $e^+e^-$, $ep$, $\bar p p$ and $pp$ colliders~\cite{Dorsner:2016wpm,Buchmuller:1986zs}, but
LQs have never been found experimentally. 
Current best experimental mass limits for the LQs set by the Large Hadron Collider (LHC) are around 1 TeV~\cite{ParticleDataGroup:2020ssz}, depending on which generation and what type. For more recent phenomenological studies of scalar LQs, see for example~\cite{Crivellin:2020ukd,Gherardi:2020qhc,Parashar:2022wrd}.

While LQs are an important topic of great interest to study, one may wonder if there are other general electroweak bosonic multiplets relevant at low energy. One can further ask a similar question as in the LQ or inert Higgs doublet~\cite{Kephart:2015oaa} case: what are the GUT origins of these more general electroweak multiplets?

In this paper, we propose to study a general spin $j$ electroweak multiplet $\phi  = \left( \phi_j, \phi_{j-1}, \cdots, \phi_{-j+1},\phi_{-j} \right)^{\rm T}$ with each of its components $\phi_m$ having mass $M_m$, where $m$ runs from $-j$ to $+j$ with $j$ either an integer or a half-integer. The possible color index carried by $\phi$ has been suppressed and color symmetry is assumed to be exact. While these general electroweak multiplets may or may not couple to the SM fermions, they must couple to the electroweak gauge bosons by construction except for the special case of $j=0$. Thus they can contribute to the electroweak gauge boson masses through loop corrections. Electroweak precision data can then put stringent constraints on their mass spectra.
In the event that the multiplet can couple to SM fermions, further constraints can be imposed by direct or indirect searches in various collider experiments. 

This paper is outlined as follows.
In Section~\ref{GUT-survey}, a GUT survey is performed to give all the electroweak irreducible representations (irreps) by decomposing the irreps from grand unification theories. These irreps not only encompass all the color-triplet/antitriplet LQs but also particles of color-sextet/antisextet and color-octet. 
In Section~\ref{STUDefinitions}, we briefly review the oblique parameters $S$, $T$ and $U$~\cite{Peskin:1991sw} and their relations to the $W$ boson mass shift $\Delta m_W$ and the $\rho$ parameter.
In Section~\ref{PiIJs},
we compute the contributions of the general electroweak multiplet to the oblique parameters. 
In Section~\ref{Analysis}, we present the constraints on the mass splittings for all the electroweak multiplets deduced from the GUT survey using the $W$ boson mass shift and the $\rho$ parameter from precision data. We consider in our analysis both the SM global fit result adapted by the Particle Data Group (PDG)~\cite{PDG:T} as well as the recent high precision $W$ boson mass measurement reported by the CDF II~\cite{CDF:2022hxs}.
We conclude in Section~\ref{conclusion}.
The interaction Lagrangian and relevant Feynman rules are given in Appendix A, while a couple of useful formulae are collected in Appendix B.

\section{GUT Survey -- Leptoquarks and General Electroweak Multiplets}
\label{GUT-survey}

LQs are particles that couple lepton to quarks. They can be vectors like the $X$ and $Y$ gauge bosons of $SU(5)$ grand unification \cite{Georgi:1974sy}, or scalars. 
Our interest will be mainly in scalar leptoquarks and other electroweak multiplets in general.
The dimension 4 couplings of scalar LQs originate in the Yukawa interactions between scalars and fermions. These scalars are a limited and well known set of fields. 
To keep our work within bounds, we will confine our study to the following gauge group chain
$$E_6\rightarrow SO(10)\rightarrow SU(5)  \rightarrow G_{\rm SM}$$ 
 where $G_{\rm SM} = SU(3)_C\times SU(2)_L \times U(1)_Y$ is the SM gauge group.
 The irreps of the larger groups $SO(10)$ and $E_6$ can all be decomposed into $SU(5)$ irreps, so we can
 begin by focusing on $SU(5)$ and discuss generalizations to larger groups later. 
 
\subsection{$SU(5)$}
\label{su5}

Fermion families of $SU(5)$ live in the representation ${\bar 5}_F+10_F$, so scalars in dimension 4 Yukawa  terms must couple to 
irreps in the products\\

\noindent
\begin{align}
\bar{5} \times \bar{5}= \overline{10}+\overline{15}\\
\bar{5} \times 10 = 5+45\\
10 \times 10 = \bar{5}+\overline{45}+\overline{50}
\end{align}

Decomposing scalar  irreps of these types (plus a few more irreps we will need later) into the standard model gauge group $G_{\rm SM}$ gives, \noindent
\begin{equation}
5  \rightarrow (3,1)_{(2)}
   +(1,2)_{(-3)} \;,
\end{equation}

\begin{equation}
15  \rightarrow (3,2)_{(-1)}+(6,1)_{(4)}+(1,3)_{(-6)} \;,
\end{equation}

\begin{equation}
10 \rightarrow \left(\bar{3},1\right)_{(4)}+(3,2)_{(-1)}+(1,1)_{(-6)} \;,  
\end{equation}

\begin{equation}
24  \rightarrow \left(\bar{3},2\right)_{(-5)}+(1,1)_{(0)}+(1,3)_{(0)}+(8,1)_{(0)}+(3,2)_{(5)} \; ,
\end{equation}

\begin{equation}
40 \rightarrow  \left(\bar{6},2\right)_{(-1)}+\left(\bar{3},1\right)_{(4)}+
   \left(\bar{3},3\right)_{(4)}+(3,2)_{(-1)}+(8,1)_{(-6)}+(1,2)_{(9)} \;,
\end{equation}

\begin{equation}
45  \rightarrow  \left(\bar{6},1\right)_{(2)}+\left(\bar{3},2\right)_{(7)}
 +\left(\bar{3},1\right)_{(-8)}+(3,1)_{(2)}+(3,3)_{(2)}+(1,2)_{(-3)}+(8,2)_{(-3)} \;,
\end{equation}
and
\begin{equation}
50 \rightarrow   \left(\bar{6},3\right)_{(2)}+\left(\bar{3},2\right)_{(7)}+(1,1)_{(12)}
   +(3,1)_{(2)}+(8,2)_{(-3)}+(6,1)_{(-8)} \; ,
\end{equation}
where
the LQs are in various $(3, 3)$, $(3,2)$,  $(3,1)$ and their conjugate irreps as summarized in \cite{ParticleDataGroup:2020ssz}.
(The group products and irrep decompositions used here and below have all been obtained with LieART \cite{Feger:2012bs,Feger:2019tvk}.
We note that the hypercharge output from LieART have been normalized to integer value as indicated by the value in the subscripts of the irrep. 
The physical hypercharge $Y$ of each irrep can be obtained by dividing the value in the subscript by 6. For example, the irrep $(8,2)_{(-3)}$ has $Y=-3/6=-1/2$ and so on.)

Besides LQs, we also obtained the following irreps:
color singlets $(1,3)$, $(1,2)$ and $(1,1)$; color-sextets/antisextets $(\bar 6,3)$, $(\bar 6,2)$ and $(6 /  \bar 6,1)$; and color-octets $(8,2)$ and $(8,1)$. Note that from the GUT survey, the allowed electroweak irreps can only reside in either the singlet, doublet or triplet, whereas the color irreps can fall into either the singlet, triplet/antitriplet, sextext/antisextet, or octet.
The doublet color-singlet (color-antisextet) $(1,2)_9$  ($(\bar 6,2)_{-1}$) can be interpreted as an elementary bilepton (biquark), according to an elegant classification of all elementary bifermions (including leptoquarks, biquarks and bileptons) based on $SU(15)$ noted recently in~\cite{Coriano:2023dfu}.

\subsection{$O(10)$}
\label{so10}

Fermion families of $SO(10)$ live in the representation  $16_F$, so scalars in dimension 4
Yukawa terms must couple to irreps in the product, 
\begin{equation}
16 \times 16 =  10+120+\overline{126} \;,
\end{equation}
while vectors couple to fermions at dimension 4 
via the gauge boson adjoint $45$.

Decomposing scalar irreps of these type into $SU(5)\times U(1)'$ and dropping the $U(1)'$  charges, assuming we are not interested in flipped models, gives

\begin{equation}
10  \rightarrow  5  +  \bar{5} \; ,
\end{equation}
\begin{equation}
120  \rightarrow   \bar{5} + \overline{45}  + \overline{10} +10 + 45 + 10 \; ,
   \end{equation}
\begin{equation}
\overline{126}  \rightarrow \bar{5} + \overline{50} + \overline{15} +1 + 45 + 10 \; .
\end{equation}

Next decomposing these irreps to the SM gauge group $G_{\rm SM}$   results in  irreps already shown  for the $SU(5)$ case above.
The vector adjoint decomposes to 
\begin{equation}
45  \rightarrow  24 + \overline{10} + 10 + 1 \; ,
\end{equation}
of $SU(5).$ For decomposition to the SM again see above.

\subsection{$E_6$}
\label{e6}

Fermion families of $E_6$ live in the representation  $27_F$, so scalars in dimension 4
Yukawa terms must couple to irreps in the product,
 \begin{equation}
27 \times 27 = \overline{27}+\overline{351}+\overline{351}'
\end{equation}
 while vectors couple at dimension 4 
to the fermions via the gauge boson adjoint $78$.
Decomposing scalar irreps of these type into $SO(10)\times U(1)''$ and dropping the  $U(1)''$ charges gives

\begin{equation}
27  \rightarrow \overline{16} + 10 + 1 \;,
\end{equation}

\begin{equation}
\overline{351}  \rightarrow \overline{16} + \overline{144} + 10 + 120 + 45 + 16 \;,
\end{equation}

\begin{equation}
\overline{351}'  \rightarrow   \overline{144} +  \overline{126} + 10 + 54 + 16 + 1 \; .
\end{equation}
We have the decomposition for all these irreps to $SU(5)$ except for the  $\overline{144}$,  $\overline{126}$ and $54$, which are
\begin{equation}
\overline{144}  \rightarrow \bar{5} + \overline{40} + \overline{15} + \overline{10} +5+ 45 + 24 \; ,
\end{equation}

\begin{equation}
\overline{126}  \rightarrow  \bar{5} + \overline{50} + \overline{15}  + \overline{10} + 1 + 45 + 10 \; ,
\end{equation}
and
\begin{equation}
54  \rightarrow   \overline{15} + 24 + 15 \; .
\end{equation}

 Finally the vector $78$ of $E_6$ decomposes to
 \begin{equation}
78  \rightarrow  1+ 16+ \overline{16} + 45  
\end{equation}
 of $SO(10)$. Further decompositions are as above.

\section{Oblique Parameters -- W Boson Mass Shift and $\rho$ Parameter}
\label{STUDefinitions}

The oblique parameters $S$, $T$, and $U$~\cite{Peskin:1991sw} represent the most important electroweak radiative corrections since they are defined by
the transverse pieces of the vacuum polarization tensors of the SM vector gauge bosons. They are process independent whereas 
the other vertex and box corrections are necessarily attached to the particles in the initial and final states in the elementary processes in high precision experiments. 
Their definitions are, defined with an overall factor of $\hat \alpha=\hat e^2/4\pi$ extracted out in front as in~\cite{Peskin:1991sw}
\begin{align}
\label{S}
\hat \alpha S & =  4 \hat s_W^2 \hat c_W^2 \left[ \Pi^\prime_{ZZ}(0)  - \frac{\hat c_W^2 - \hat s_W^2}{\hat s_W \hat c_W} \Pi^\prime_{Z\gamma}(0)  - \Pi^\prime_{\gamma\gamma}(0) \right],\;\;\;\;\; \\
\label{T}
\hat \alpha T & =  \frac{\Pi_{WW}(0)}{m_W^2} - \frac{\Pi_{ZZ}(0)}{m_Z^2} \; ,  \\ 
\label{U}
\hat \alpha U & =  4 \hat s_W^2  \big[ \Pi^\prime_{WW}(0)  - \hat c_W^2 \Pi^\prime_{ZZ}(0)  - 2 \hat s_W \hat c_W \Pi^\prime_{Z\gamma}(0)  - \hat s_W^2 \Pi^\prime_{\gamma\gamma}(0) \big] \; .
\end{align}
Here the hat quantities $\hat c_W$, $\hat s_W$ and $\hat \alpha$ are understood to be evaluated at the $Z$ mass pole. 
In Eqs.~(\ref{S}), (\ref{T}) and (\ref{U}), as mentioned above, $\Pi_{IJ}(0)$ is the transverse part of vacuum polarization tensor $ \Pi_{IJ}^{\mu\nu}(q)$ evaluated at $q^2=0$ for gauge bosons $I$ and $J$,
\begin{equation}
 \Pi_{IJ}^{\mu\nu}(q) =  \Pi_{IJ} (q^2) g^{\mu\nu} - \Delta_{IJ} (q^2) q^\mu q^\nu  \; .  
\end{equation}
The second term needs no concern to us since at  high energy experiments like LEP I and II where electroweak precision 
measurements were carried out, $q^\mu$ will dot into the currents formed by helicity spinors of light leptons 
and will give nil results. And $\Pi^\prime_{IJ}(0) = d \Pi_{IJ}(q^2)/dq^2 \vert_{q^2=0}$.

As alluded to in the introduction, the general electroweak irreps obtained from the GUT survey in the previous section may or may not couple to the SM fermions, but they are necessarily coupled to the SM gauge bosons as long as they are not electroweak singlets. At one loop these multiplets will give contributions to the vacuum polarization amplitudes $\Pi_{IJ}$ and hence the oblique parameters too. Electroweak precision data from global fit of the oblique parameters can provide stringent constraints on the mass splittings within each irrep. For those multiplets which are electroweak singlets, their oblique parameters vanish. 
Thus we will not consider 
$(1,1)$, $(3 / \bar 3,1)$, $(6 / \bar 6,1)$ and $(8,1)$ in this work.

\subsection{$W$ Boson Mass}
\label{MW}

Recently ATLAS~\cite{ATLAS-CONF-2023-004}, using a more modern set of parton distribution functions of the proton (CT18 set~\cite{Hou:2019efy}), has reported a new precision measurement of the $W$ boson mass
\begin{equation}
\label{mw-atlas}
m_W ({\rm ATLAS} ) = 80,360 \pm 16 \; {\rm MeV}/c^2 \; ,
\end{equation}
which has a 10 MeV decrease in the central value and a 3 MeV reduction in the total uncertainty as compared with the previous result from 2017~\cite{ATLAS:2017rzl}.
No deviation is observed as compared with 
the SM prediction from electroweak global fit~\cite{deBlas:2021wap}
\begin{equation}
\label{mw-globalfits}
m_W ({\rm SM - \; Global \, Fit}) = 80,359.1 \pm 5.2 \; {\rm MeV/}c^2 \; .
\end{equation}
On the other hand, the CDF II result~\cite{CDF:2022hxs} a year ago is
\begin{equation}
\label{mw-cdf}
    m_W ({\rm CDF \, II} )= 80,433.5 \pm 9.4 \; {\rm MeV}/c^2 \; ,
\end{equation}
which is $\sim 7 \sigma$ away from the SM global fit in Eq.~(\ref{mw-globalfits}).
Besides the impacts of these new measurements to the electroweak precision global fit~\cite{Lu:2022bgw,deBlas:2022hdk,Asadi:2022xiy}, 
it could entail new physics beyond the SM (bSM). 

While combined result of the measurements from  LEP, Tevatron and ATLAS are still lacking, 
pending evaluation of uncertainty correlations~\cite{CDF:2022hxs,2134139}, 
bSM enthusiasts have already offering 
numerous new physics interpretations~\cite{Fan:2022dck,Cheung:2022zsb,Ahn:2022xax,Kanemura:2022ahw,Du:2022fqv,Cheng:2022aau,Lee:2022gyf,Abouabid:2022lpg,Chen:2022ocr,Evans:2022dgq,Senjanovic:2022zwy,Bahl:2022gqg,Barger:2022wih,Tran:2022yrh,Tran:2022cwh,Wu:2022uwk} of the CDF II result.

The $W$ boson mass shift is related to the oblique parameters according to~\cite{Peskin:1991sw}
\begin{align}
\label{wmassshift}
\frac{ \Delta m_W^2 }{m_Z^2}  & =  \hat \alpha \frac{\hat c_W^2 }{\hat c_W^2 - \hat s_W^2} \left[ -\frac{S}{2} + \hat c_W^2 T 
+ \frac{\hat c_W^2 - \hat s_W^2}{4 \hat s_W^2} U  \right] . \;\;\;
\end{align}

In order to compare with the difference of the central values of the ATLAS (\ref{mw-atlas}) (CDF II~(\ref{mw-cdf})) result and  the global fit number (\ref{mw-globalfits}), namely, 
$\Delta m_W \approx$ 1 (75) MeV, 
we use $\Delta m_W \approx  \left( \Delta m_W^2 + m_W^2 \right)^{1/2} - m_W$ and $\Delta m_W^2$ by (\ref{wmassshift}).
We set $\hat \alpha = 1/127.951$, $m_W = 80.3591$ GeV, $m_Z = 91.1876$ GeV and $\hat c_W = m_W/m_Z=0.88125$ in our numerical work.

\subsection{$\rho$ Parameter}
\label{rhoParameter}

As is well known, the $\rho$ parameter is defined as 
\begin{equation}
    \rho = \frac{M_{W^+}^2 - M_{W^3}^2} {M_W^2} \; ,
\end{equation}
which measures the mass splitting of the weak gauge bosons in the isovector representation of  $SU(2)_L$.
In the SM, $\rho = 1$ at tree level. New bSM particles that give rise to different contributions to the $W^+$ and $W^3$ masses at one-loop can provide a shift to the $\rho$ parameter.
The change of the $\rho$ parameter (with respect to the SM tree plus high-order contributions) is related to the oblique parameter $T$ from bSM as
\begin{equation}
\label{rhoTRelation}
    \Delta \rho = \hat \alpha T \; .
\end{equation}

The SM global fit prediction, as adapted by the PDG~\cite{PDG:T}, gives the $2\sigma$ contours of the oblique parameters in the range of 
\begin{align}
S & = -0.02 \pm 0.10 \; ,\\ 
T & = 0.03 \pm 0.12 \; , \\
U & = 0.01 \pm 0.11 \; ,
\end{align}
which contains the SM (with $(S,T,U)=(0,0,0)$).
On the other hand, according to the analysis in~\cite{Asadi:2022xiy}, the SM  is strongly disfavored when the new CDF II $W$ boson mass measurement is included in their global fit. In particular, $T$ is now positive in the range of $0.12 < T < 0.42$.

\section{$\Pi_{IJ}(q^2)$ for the general scalar electroweak multiplets}
\label{PiIJs}

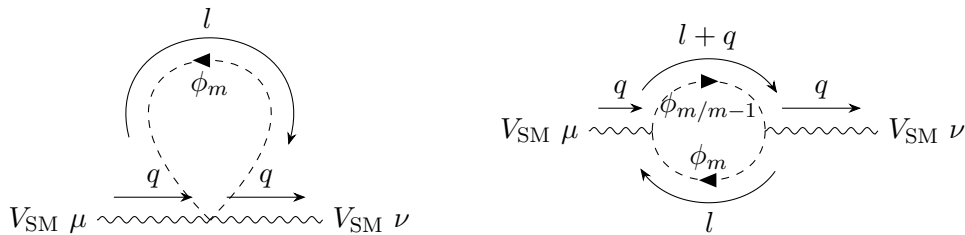
\begin{figure}[H]
       \centering
\begin{tikzpicture}
\begin{feynman}
\vertex (b) ;
\vertex [left=of b] (d) {\(V_{\rm SM}~\mu\)};
\vertex [right=of b] (e) {\(V_{\rm SM}~\nu\)};
\diagram* {
b -- [out=135, in=45, loop, anti charged scalar,  momentum=$l$, min distance=4cm, edge label'={\(\phi_m \)}] b,
(d) -- [boson, momentum = $q$] (b),
(b) -- [boson, momentum = $q$] (e),
};
\end{feynman}
\end{tikzpicture}
\qquad
\begin{tikzpicture}
\begin{feynman}
\vertex (a){\(V_{\rm SM}~\mu\)};
\vertex [right=of a] (b);
\vertex [right=of b] (c) ;
\vertex [right=of c] (d) {\(V_{\rm SM}~\nu\)};
\diagram* {
(a) -- [boson, momentum = $q$] (b),
(b) -- [charged scalar, half left, momentum=$l+q$, edge label'={\(\phi_{m/m-1} \)} ] (c),
(c) -- [charged scalar, half left, momentum=$l$, edge label'={\(\phi_{m} \)} ] (b),
(c) -- [boson, momentum = $q$ ] (d),
};
\end{feynman}
\end{tikzpicture}
\caption{\label{bubble-loop} One-loop Feynman diagrams for vacuum polarization tensors of SM vector bosons with $V_{\rm SM}$ stands for $\gamma$, $Z$ or $W^-$.
}
\end{figure}

We now present the vacuum polarization amplitudes $\Pi_{IJ}(q^2)$ for a general scalar electroweak multiplet at one-loop. The interaction Lagrangian and relevant Feynman rules are given in Appendix A. The relevant Feynman diagrams are depicted in Fig.~\ref{bubble-loop}. 

For the charged gauge boson $I=J=W^-$, we have
\begin{align}
\label{PiWWq}
\Pi_{WW} (q^2) & =  \frac{1}{8 \pi} N_C \left( \frac{ \hat \alpha }{ \hat s_W^2 } \right)
\left\{ 
\sum_{m=-j}^{+j} \langle j \, m \vert \left\{t^+ \, ,t^- \right\} \vert j \, m \rangle M_m^2 \log \left( \frac{M_m^2}{\mu^2} \right) \right. \nonumber \\
& +  \left. \sum^{+j}_{m=-j+1} \langle j \, m-1 \vert t^- \vert j \, m \rangle ^2 \left[ \frac{1}{3} q^2 \left( E +1 \right) - 2 B \left( M_m, M_{m-1}, q^2 \right) \right]
\right\} \; .
\end{align}
Here $N_C$ is the dimension of color representation of the general electroweak multiplet. For instance, for the
leptoquarks in either the fundamental irrep $3$ or $\bar 3$ of $SU(3)_C$, $N_C = 3$, while $N_C = 6$ and 8 for the sextet or antisextet and octet respectively.
In Eq.~(\ref{PiWWq}), 
\begin{align}
E & = \frac{2}{\epsilon} + \log 4 \pi - \gamma_E \; ,\\
\langle j \, m \vert \left\{t^+ \, ,t^- \right\} \vert j \, m \rangle & =  2 \left( j \left( j + 1 \right) - m^2 \right) \; ,\\
 \langle j \, m-1 \vert t^- \vert j \, m \rangle & = \sqrt{ j \left( j + 1 \right) - m \left( m - 1 \right) } \; ,
\end{align}
and $B(M_1,M_2,q^2)$ is the integral defined in Appendix B.

For the neutral gauge bosons $I, J=\gamma$ or $Z$, one obtains
\begin{align}
\Pi_{\gamma\gamma} (q^2) & =  
\frac{1}{4 \pi} N_C \hat \alpha
\sum_{m=-j}^{+j} Q_m^2 \left[  \frac{1}{3} q^2 \left( E +1 \right) + 2 M_m^2 \log \left( \frac{M_m^2}{\mu^2} \right)  - 2 B \left( M_m, M_m, q^2 \right)
\right] \; ,
\end{align}
where $Q_m = m + Y$ with $Y$ the hypercharge; 
\begin{align}
\Pi_{ZZ} (q^2) & =  \frac{1}{4 \pi} N_C \left( \frac{ \hat \alpha  }{\hat s^2_W \hat c^2_W} \right) \nonumber \\
& \times  \sum_{m=-j}^{+j} \left( m - \hat s_W^2 Q_m \right)^2 \left[  \frac{1}{3} q^2 \left( E +1 \right) + 2 M_m^2 \log \left( \frac{M_m^2}{\mu^2} \right)  - 2 B \left( M_m, M_m, q^2 \right)
\right] \; ,
\end{align}
and 
\begin{align}
\Pi_{\gamma Z} (q^2) & = 
 \frac{1}{4 \pi} N_C \left( \frac{ \hat \alpha  }{\hat s_W \hat c_W} \right) \nonumber \\
& \times  \sum_{m=-j}^{+j} Q_m \left( m - \hat s_W^2 Q_m \right) \left[  \frac{1}{3} q^2 \left( E +1 \right) + 2 M_m^2 \log \left( \frac{M_m^2}{\mu^2} \right)  - 2 B \left( M_m, M_m, q^2 \right)
\right] \; .
\end{align}

Note that $\Pi_{\gamma\gamma}(0) = \Pi_{\gamma Z}(0) = \Pi_{ZZ}(0) = 0$! 
So the $T$ parameter is contributed entirely from $\Pi_{WW}(0)$. One can easily show that unless for a degenerate multiplet, $\Pi_{WW}(0)$ is in general  non-vanishing.
Using the above expressions of the vacuum polarization amplitudes and the simple formulas related to the $B(M_1,M_2,q^2)$ integral in Appendix B, 
it is straightforward to demonstrate all the $1/\epsilon$ pole terms are cancelled in the oblique parameters.
The finite results for the $S$, $T$ and $U$ oblique parameters can be written down in compact forms as
\begin{align}
S & = - \frac{1}{3 \pi} Y N_C 
\sum_{m=-j}^{+j} m 
\log \frac{M_m^2}{\mu^2} \; ,
\end{align}
\begin{align}
T & = \frac{1}{8 \pi \hat s_W^2} N_C \left[ 
 \sum_{m=-j}^{+j} \langle j \, m \vert \left\{ t^+ , t^- \right\} \vert j \, m \rangle \frac{M_m^2}{m_W^2} \log \frac{M_m^2}{\mu^2}  \right. \nonumber \\
& \qquad \qquad \qquad \left.  - 2  \sum^{+j}_{m=-j+1} \langle j \, m-1 \vert t^- \vert j \, m \rangle ^2 \frac{B\left(M_m,M_{m-1},0\right)}{m_W^2}
\right] \; ,
\end{align}
and
\begin{align}
U & =  \frac{1}{\pi} N_C  \Biggl[
2 \sum_{m=-j}^{+j} m^2  
B^\prime \left( M_m, M_m, 0 \right) - \sum^{+j}_{m=-j+1}  \langle j \, m-1 \vert t^- \vert j \, m \rangle ^2 B^\prime \left( M_m, M_{m-1}, 0 \right) \Biggr] \; .
\end{align}
Since all the divergences are cancelled, the $\mu^2$ dependence (hidden in $B$ and $B^\prime$) in the oblique parameters are  faked. 
They are cancelled in the same manner. This also implies that for a completely degenerate multiplet, the oblique parameters vanish as well, which can be checked
explicitly using the familiar identities encountered in quantum mechanics:
\bea
\label{Identity1}
 \sum_{m=-j}^{+j} m & = & 0 \; , \\
 \label{Identity2}
 \sum_{m=-j}^{+j} \langle j \, m \vert \left\{ t^+ , t^- \right\} \vert j \, m \rangle & = & 2 \sum^{+j}_{m=-j+1} \langle j \, m-1 \vert t^- \vert j \, m \rangle ^2  = 4  \sum_{m=-j}^{+j} m^2 \; , \nonumber \\
 & = & \frac{4}{3} \, j \left( j+1 \right) \left( 2j+1 \right) \; . 
\eea
It is interested to note that the hypercharge $Y$ only enters in the $S$ oblique parameter. 

We reproduced the oblique parameters computed in \cite{Lavoura:1993nq} and note that our results of $S$ and $T$ are consistent with a similar calculation in~\cite{Wu:2022uwk}. Oblique parameters within leptoquark models were also investigated in \cite{Crivellin:2020ukd}.

\section{Analysis}
\label{Analysis}

The oblique parameters computed in previous section are for a general electroweak multiplet and can apply to any electroweak irrep that we deduced from the GUT survey in Section~\ref{GUT-survey}. In this section, we will use the precision measurements of $W$ boson mass shift and $\rho$ parameter discussed in Subsections~\ref{MW} and~\ref{rhoParameter} respectively to put constraints on the mass splittings for each of these electroweak irreps.
We will separate the two cases of electroweak doublets and triplets for discussions.

\subsection{Scalar Electroweak Doublets}
\label{doublets}

\begin{figure}[H]
        \centering
\includegraphics[width=0.45\textwidth]{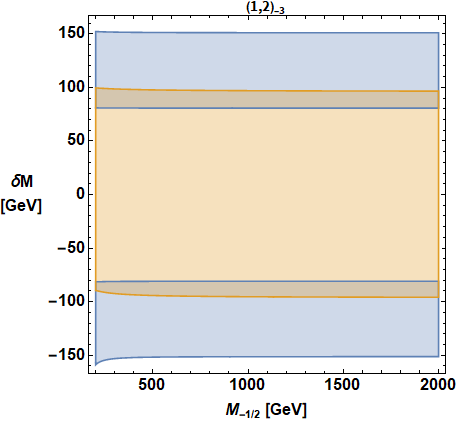}
\quad
\includegraphics[width=0.45\textwidth]{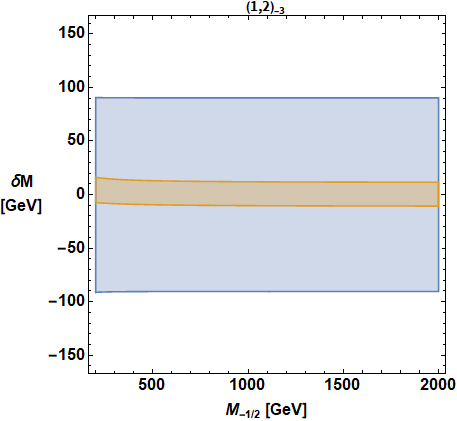}
\caption{\label{fig:doublet-colorsinglet}
Restrictions on the mass $M_{-1/2}$ of $\phi_{-1/2}$ and mass splitting $\delta M = M_{1/2} - M_{-1/2}$ between $\phi_{1/2}$ and $\phi_{-1/2}$ for the electroweak doublet representation	$(1,2)_{-3}$. Left (right) column is for the CDF II data (the SM global fit prediction from PDG). Blue (orange) regions indicating the constraints from the $\rho$ parameter (the $W$ boson mass shift).} 
\end{figure}

For a scalar doublet $\phi$ where $j = \frac{1}{2}$, Eqs.~(\ref{wmassshift}) and (\ref{rhoTRelation}) depend only on two parameters: the mass $M_{-1/2}$ of $\phi_{-1/2}$ and the mass splitting $\delta M = M_{1/2} - M_{-1/2}$ between $\phi_{1/2}$ and $\phi_{-1/2}$. According to our GUT survey, $(1,2)_{-3}$, $(1,2)_{9}$, $(3,2)_{-1}$, $(\bar 3,2)_{-5}$, $(3,2)_{5}$, $(\bar 3,2)_{7}$, $(\bar 6,2)_{-1}$ and $(8,2)_{-3}$ are the eight distinctive electroweak doublet representations. For each representation, we plot the allowed regions of $M_{-1/2}$ and $\delta M$ based on the global fit~\cite{Asadi:2022xiy} including the CDF II $W$ boson measurement which gives $0.12 < T < 0.42$ and the corresponding $\Delta m_W\leq 75$ MeV (See the plots on the left columns of Figs.~\ref{fig:doublet-colorsinglet}). We also made plots for the allowed regions of $M_{-1/2}$ and $\delta M$ based on the SM global fit adapted by the PDG~\cite{PDG:T} which prefers $-0.09 < T < 0.15$~\footnote{Since the numerical values of $T$ for all the multiplets we consider are always positive, we impose a smaller range of constraint 
$0 < T < 0.15$ in our numerical work in what follows.} and the corresponding $\Delta m_W\leq 1$ MeV (See the plots on the right columns of Figs.~\ref{fig:doublet-colorsinglet}). 

In Fig.~\ref{fig:doublet-colorsinglet}, we present the restrictions on the mass $M_{-1/2}$ of $\phi_{-1/2}$ and mass splitting $\delta M = M_{1/2} - M_{-1/2}$ between $\phi_{1/2}$ and $\phi_{-1/2}$ for the color-singlet representations	$(1,2)_{-3}$ and $(1,2)_{9}$. 
The blue regions of the plots on the left side give the allowed values of $M_{-1/2}$ and $\delta M$ satisfying $0.12< T < 0.42$ (CDF II), while the orange regions of the plots on the left side indicate the allowed values of $M_{-1/2}$ and $\delta M$ satisfying $\Delta m_W\leq 75$ MeV. The blue regions of the plots on the right side give the values of $M_{-1/2}$ and $\delta M$ when the parameter $0 < T < 0.15$ (PDG), while the orange regions of the plots on the right side indicate where are the values of $M_{-1/2}$ and $\delta M$ distributed when $\Delta m_W\leq 1$ MeV.

The plots of the other seven distinctive electroweak doublet representations $(1,2)_{9}$, $(3,2)_{-1}$, $(\bar 3,2)_{-5}$, $(3,2)_{5}$, $(\bar 3,2)_{7}$, $(\bar 6,2)_{-1}$ and $(8,2)_{-3}$ look virtually the same by eye due to very weak $M_{-1/2}$ dependencies, hence we hesitate to display them here despite the fact that they are not identical.
Detailed constraints of the doublet leptoquark $(3,2)_{-1}$ (as well as the singlet leptoquark $(\bar 3, 1)_2$) from various experiments can be found in a recent work~\cite{Parashar:2022wrd}.
In addition, a detailed collider phenomenological study for the color-octet $(8,2)_{-3}$ was done some time ago by Manohar and Wise~\cite{Manohar:2006ga}.

\subsection{Scalar Electroweak Triplets}
\label{triplets}

\begin{figure}[b!]
        \centering
\includegraphics[width=0.41\textwidth]{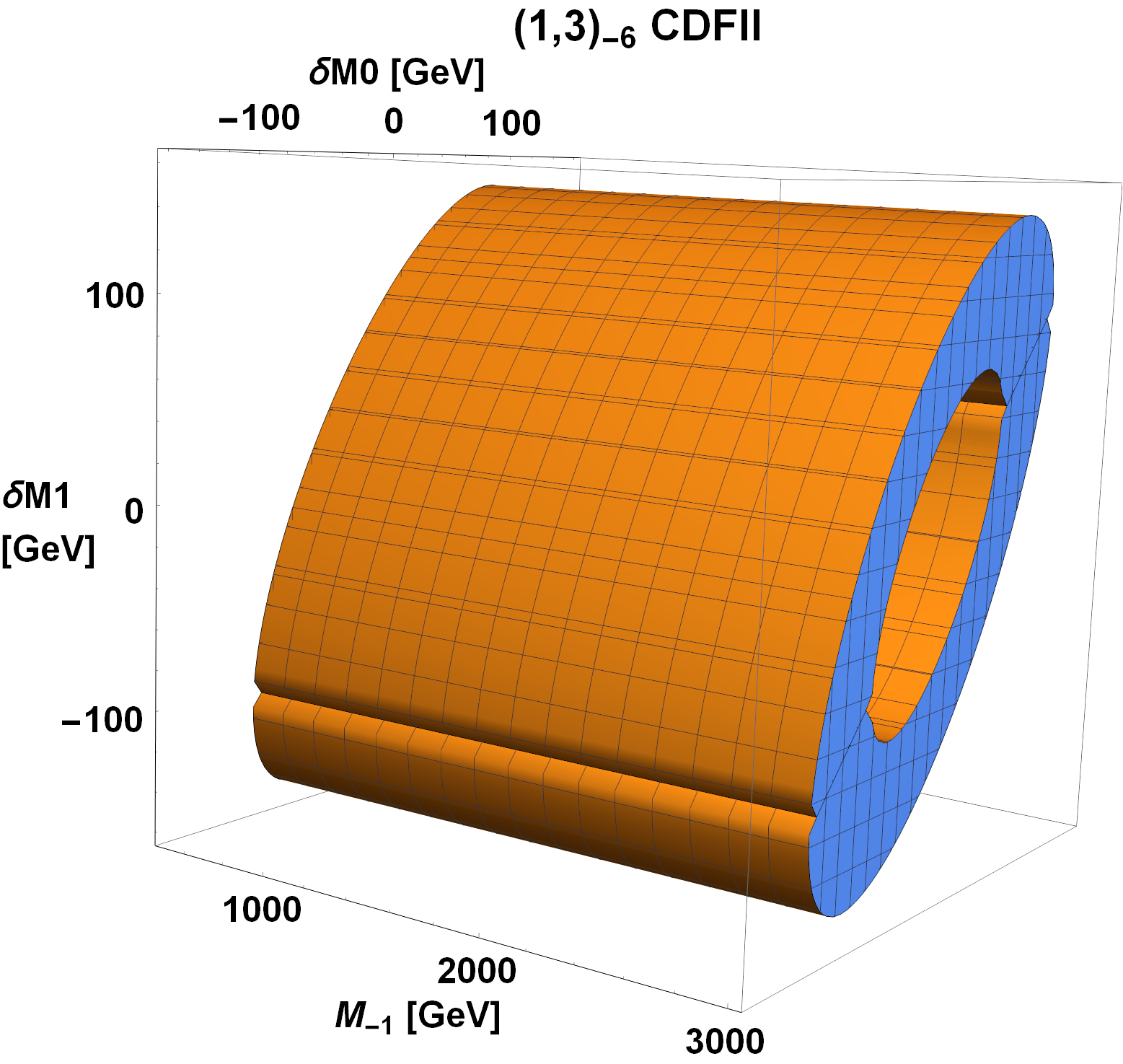}
\quad
\includegraphics[width=0.41\textwidth]{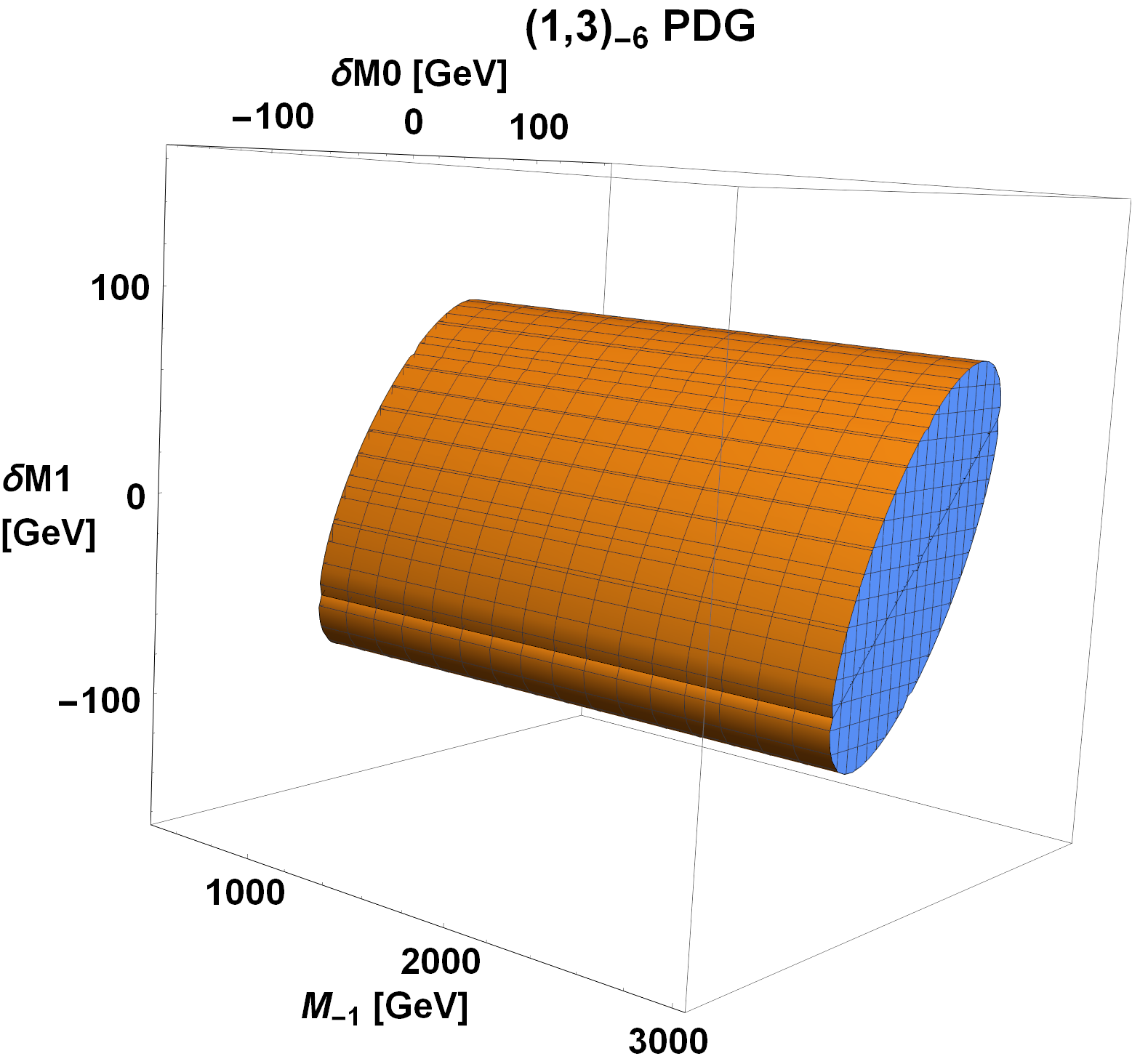}
\\~~\\
\includegraphics[width=0.41\textwidth]{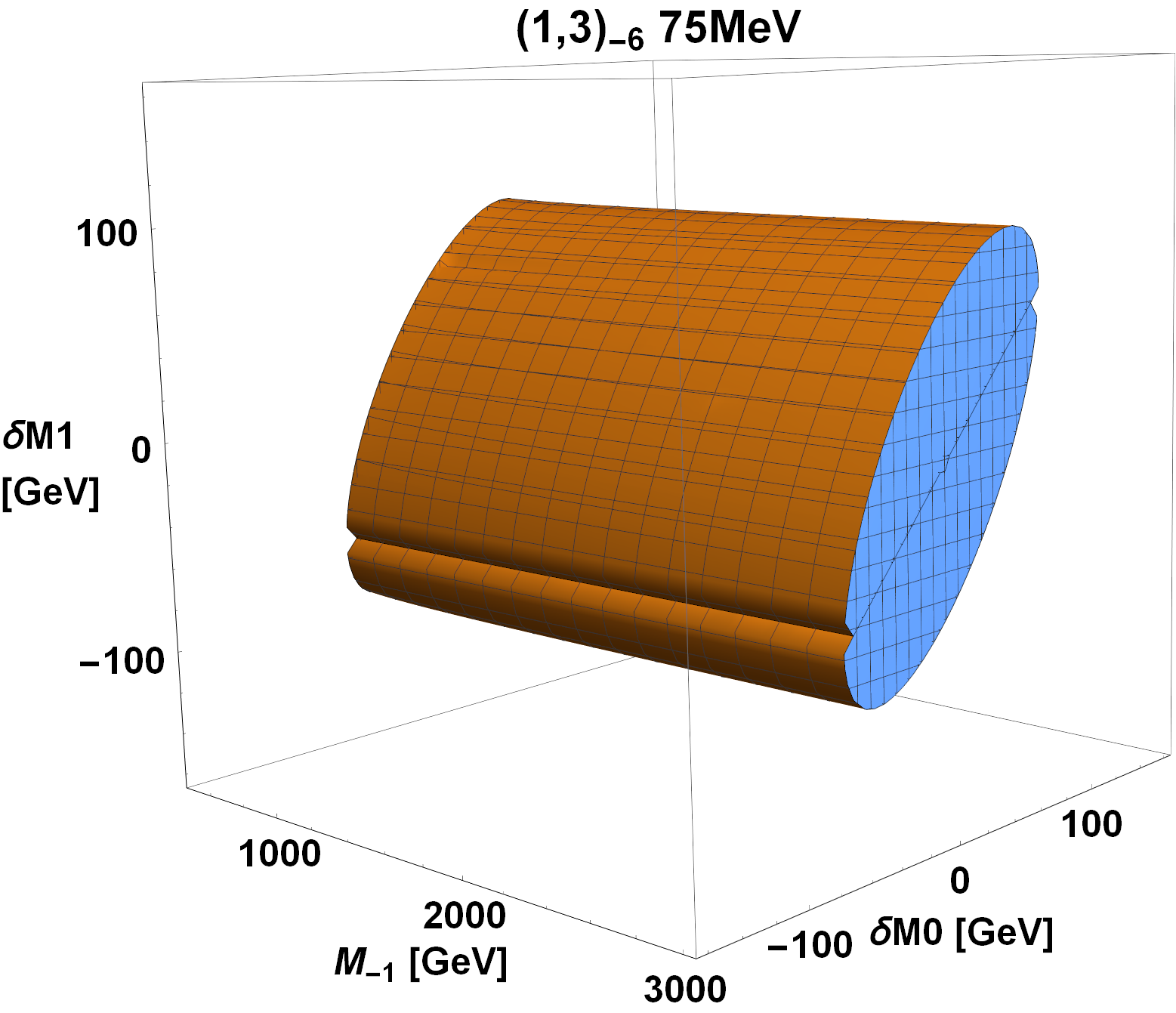}
\quad
\includegraphics[width=0.41\textwidth]{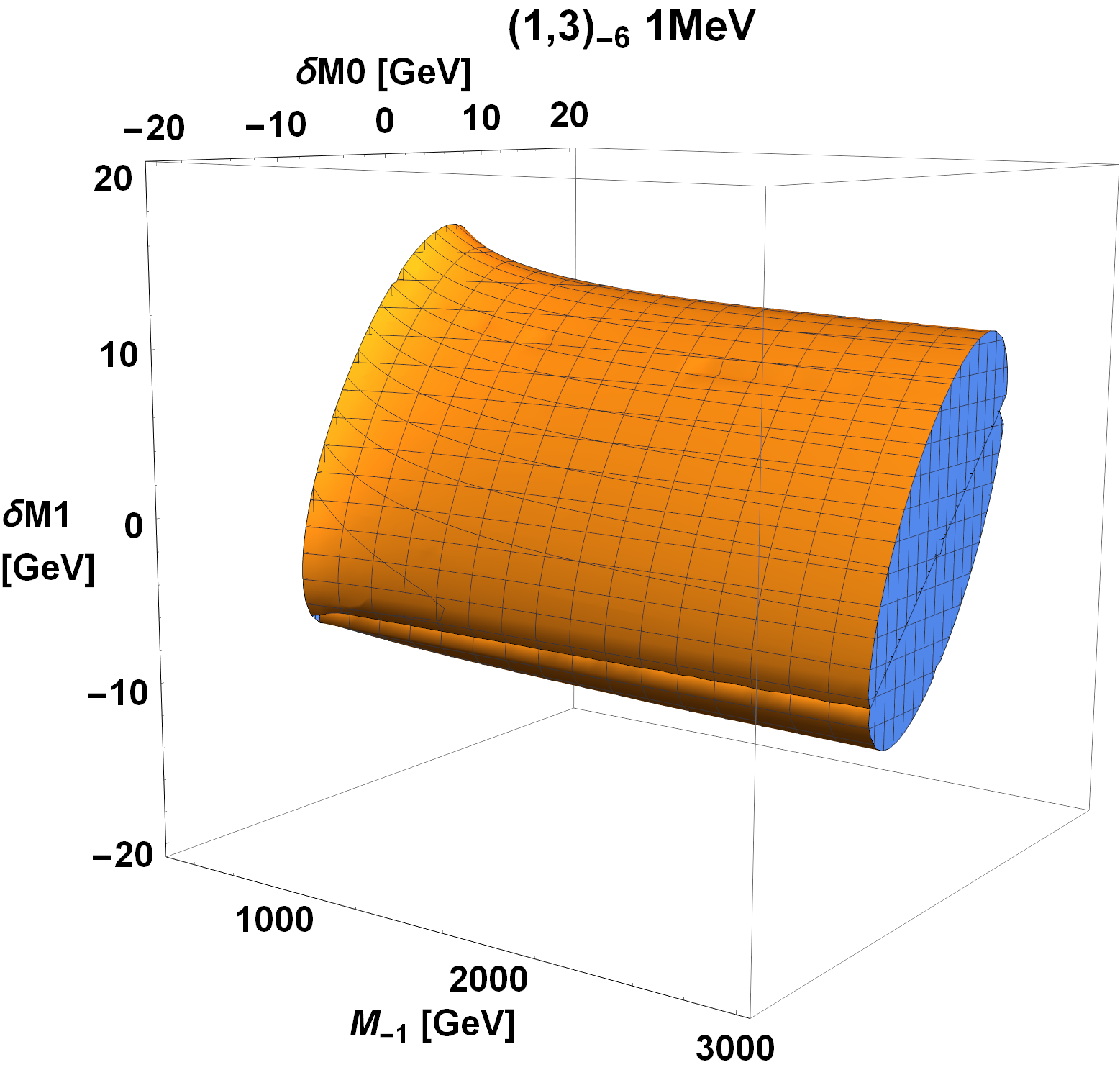}
\caption{\label{fig:1,3,-6}
Restrictions on the mass $M_{-1}$ of $\phi_{-1}$, the mass splittings $\delta M0$ and $\delta M1$ for color-singlet electroweak triplet $(1,3)_{-6}$ irrep. The top left (bottom left) panel is for $0.12 < T < 0.42$ ($\Delta m_W\leq 75$ MeV) due to the CDF II result, while the top right (bottom right) panel 
is for $0 < T < 0.15$ ($\Delta m_W\leq 1$ MeV) from SM global fit adapted by PDG.
} 
\end{figure}

\begin{figure}[H]
        \centering
\includegraphics[width=0.45\textwidth]{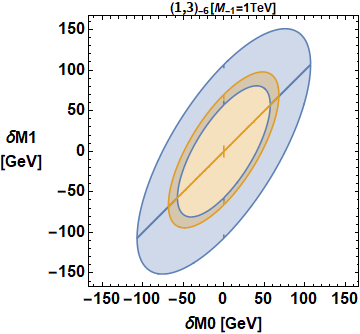}
\quad
\includegraphics[width=0.45\textwidth]{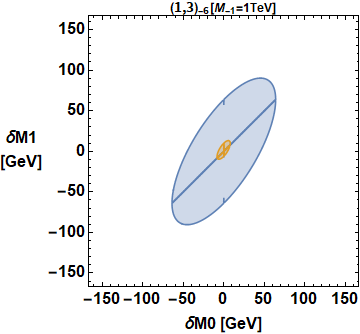}
\\~~\\
\includegraphics[width=0.45\textwidth]{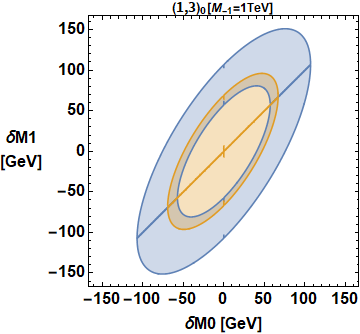}
\quad
\includegraphics[width=0.45\textwidth]{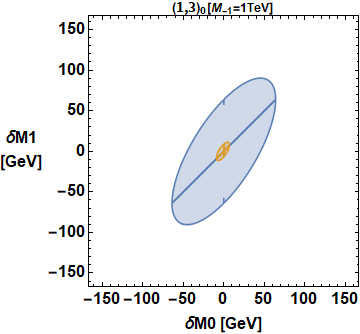}
\caption{\label{fig:triplet-colorsinglet}
Restrictions on the mass splittings $\delta M0$ and $\delta M1$ with $M_{-1}=1$ TeV for the color-singlet electroweak triplets $(1,3)_{-6}$ (top panels) and $(1,3)_{0}$ (bottom panels). The blue and orange regions in the plots are for the $T$ parameter and $\Delta m_W$ respectively. Left column is for CDF II result with $0.12 < T < 0.42$ and $\Delta m_W\leq 75$ MeV, while the right column is for SM global fit adapted by PDG with $0 < T < 0.15$ and $\Delta m_W \leq 1$ MeV.} 
\end{figure}

\begin{figure}[H]
        \centering
\includegraphics[width=0.45\textwidth]{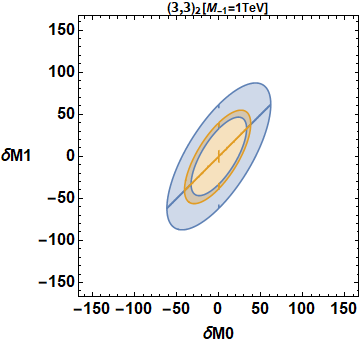}
\quad
\includegraphics[width=0.45\textwidth]{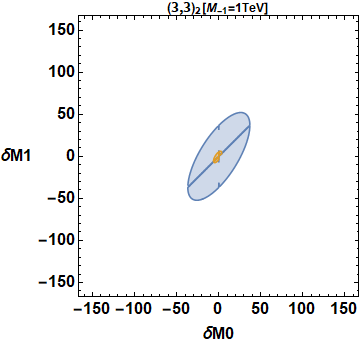}
\\~~\\
\includegraphics[width=0.45\textwidth]{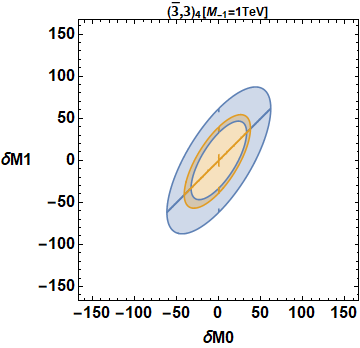}
\quad
\includegraphics[width=0.45\textwidth]{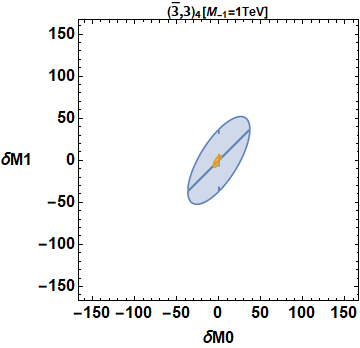}
\caption{\label{fig:triplet-colortriplet}
Similar to Fig.~\ref{fig:triplet-colorsinglet} but for the color-triplet electroweak triplets
$(3,3)_{2}$ and $(\bar 3,3)_{4}$.
} 
\end{figure}

\begin{figure}[H]
       \centering
\includegraphics[width=0.45\textwidth]{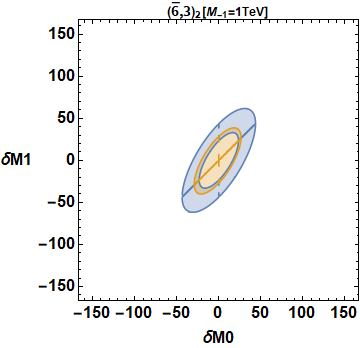}
\quad
\includegraphics[width=0.45\textwidth]{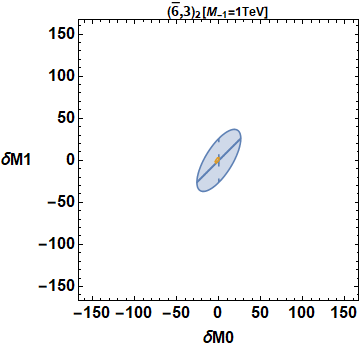}
\caption{\label{fig:triplet-colorsextet}
Similar to Figs.~\ref{fig:triplet-colorsinglet} and \ref{fig:triplet-colortriplet} but for the color-antisextet electroweak triplet$(\bar 6,3)_{2}$.
}
\end{figure}

For a scalar triplet $\phi$ with $j=1$, Eqs.~(\ref{wmassshift}) and (\ref{rhoTRelation}) now depend on three parameters: the mass $M_{-1}$ of $\phi_{-1}$, the mass splitting $\delta M0 = M_{0} - M_{-1}$ between $\phi_{0}$ and $\phi_{-1}$, and the mass splitting $\delta M1 = M_{1} - M_{-1}$ between $\phi_{1}$ and $\phi_{-1}$. According to our GUT survey, $(1,3)_{-6}$, $(1,3)_{0}$, $(3,3)_{2}$, $(\bar 3,3)_{4}$ and $(\bar 6,3)_{2}$ are the five distinctive electroweak triplet representations. Take $(1,3)_{-6}$ as an example. In the top left and bottom left (top right and bottom right) panels in the 3D plots in Fig.~\ref{fig:1,3,-6}, we show the allowed regions for $M_{-1}$, $\delta M0$ and $\delta M1$ with $0.12 < T < 0.42$ and $\Delta m_W \leq 75$ MeV respectively due to the CDF II result ($0 < T < 0.15$ and $\Delta m_W \leq 1$ MeV respectively due to the SM global fit adapted by PDG). Since there isn't much variation in both $\delta M0$ and $\delta M1$ as $M_{-1}$ increases from 500 GeV to 3 TeV, we slice a cross section of the 3D plots at $M_{-1}=1$ TeV and show restrictions on $\delta M0$ and $\delta M1$ in Figs.~\ref{fig:triplet-colorsinglet}, \ref{fig:triplet-colortriplet} and \ref{fig:triplet-colorsextet} for the aforementioned electroweak triplets.

In Fig.~\ref{fig:triplet-colorsinglet}, we show the constraints on the mass splitting parameters 
$\delta M0$ and $\delta M1$ with $M_{-1}=1$ TeV for the color-singlet $(1,3)_{-6}$ (top two panels) and $(1,3)_0$ (bottom two panels). The left column is for the CDF II with $0.12 < T < 0.42$ (blue region) and $\Delta m_W \leq 75$ MeV (orange region), while the right column is for the PDG with $0 < T < 0.15$ (blue region) and $\Delta m_W \leq 1$ MeV (orange region).

Fig.~\ref{fig:triplet-colortriplet} is similar to Fig.~\ref{fig:triplet-colorsinglet} but for the two color-triplets  $(3,3)_2$ and $(\bar 3, 3)_4$.

Similar to Figs.~\ref{fig:triplet-colorsinglet} and~\ref{fig:triplet-colortriplet}, Fig.~\ref{fig:triplet-colorsextet} show the constraints on the mass splitting parameters for the color-antisextet $(\bar 6,3)_2$. 

We end this section by summarizing our findings as follows.

\begin{itemize}

\item
For both electroweak doublets and triplets, when the masses $M_{-1/2}$ and $M_{-1}$ of the ``lowest'' components $\phi_{-1/2}$ and $\phi_{-1}$ respectively change from 500 GeV to 3 TeV, there aren't much variations for the mass splittings $\delta M$ for the doublets and $\delta M0$ and $\delta M1$ for the triplets. In other words, the mass splittings have almost no dependence on the overall mass scale of the multiplets in this mass range. Most of the variations are in the lower mass scale which may not be favorable or even excluded readily for these general electroweak multiplets, given the best limits on the leptoquark masses from LHC have reached $\sim 1$ TeV. It's reasonable to infer that the mass splittings between multiplet components depend very weakly on the mass $M_m$ each component $\phi_m$ carries.

\item
We can understand analytically why the dependencies on the overall mass scale are very weak across the plots shown above. It is sufficient to consider the doublet case of $j={1\over2}$. Let $M = M_{-\frac{1}{2}}$ and $\Delta M^2  = M^2_{1\over2} - M^2_{-{1\over2}}$. The latter is related to the linear mass difference $\delta M = M_{1\over2} - M_{-{1\over2}}$ used in the plots as $\Delta M^2 \approx 2 M \delta M$.
We expand the oblique parameters in terms of the dimensionless ratio $\Delta M^2 /M^2$ (or $\delta M /M$) assuming to be small here. We also assume $M \gg m_W$. One finds
\begin{flalign}
S_{1\over2} ~~ = & ~~  - \frac{1}{6 \pi} Y N_C \log {  M_{1\over2}^2 \over M_{-{1\over2}}^2} \; , \nonumber \\
\approx & ~~ - \frac{1}{6 \pi} Y N_C  {\Delta M^2 \over M^2} \left[ 1
+ {\mathcal O} \left( {\Delta M^2 \over M^2 } \right) \right] \; , \\
\approx & ~~ - \frac{1}{3 \pi} Y N_C  {\delta M \over M} \left[ 1
+ {\mathcal O} \left( {\delta M \over M } \right) \right] \; .
\end{flalign}
\begin{flalign}\nonumber  
 T_{1\over2} ~~ = & ~~ \frac{1}{8 \pi \hat s_W^2} N_C \left[ 
 {  M_{1\over2}^2 + M_{-{1\over2}}^2 \over {2 m_W^2} } -  { M_{1\over2}^2 M_{-{1\over2}}^2 \over  {({M_{1\over2}^2 - M_{-{1\over2}}^2}) m_W^2}  } \log { M_{1\over2}^2 \over M_{-{1\over2}}^2} \right] ~ \; ,  \nonumber \\
 ~~ \approx & ~~ 
\frac{1}{48 \pi \hat s_W^2} N_C \frac{(\Delta M^2)^2}{m_W^2 M^2 } \left[ 1 + {\mathcal O} \left( {\Delta M^2 \over M^2 } \right) \right] \; , \\
 ~~ \approx & ~~ 
\frac{1}{12 \pi \hat s_W^2} N_C \frac{\delta M^2}{m_W^2} \left[ 1 + {\mathcal O} \left( {\delta M \over M } \right) \right] \; .
\end{flalign}
\begin{flalign}  
U_{1\over2} ~~ = & ~~  \frac{N_C}{\pi} \left[ - \frac{5}{36} +
  { M_{-{1\over2}}^2 M_{1\over2}^2 \over 3( M_{1\over2}^2 - M_{-{1\over2}}^2 )^2 }
+  { M_{1\over2}^6 + M_{-{1\over2}}^6 - 3 M_{-{1\over2}}^2 M_{1\over2}^2 ( M_{1\over2}^2 +  M_{-{1\over2}}^2 )   \over 12 ( M_{1\over2}^2 - M_{-{1\over2}}^2 )^3 }\log {M_{1\over2}^2 \over M_{-{1\over2}}^2}
 \right] \; , \nonumber \\
     ~~ \approx & ~~ \frac{1}{60 \pi} N_C \frac{(\Delta M^2)^2}{M^4} \left[ 1 + {\mathcal O} \left( {\Delta M^2 \over M^2 } \right) \right] \; , \\
 ~~ \approx & ~~ \frac{1}{15 \pi} N_C \frac{\delta M^2}{M^2} \left[ 1 + {\mathcal O} \left( {\delta M \over M } \right) \right] \; .
\end{flalign}
Note that $S_{\frac{1}{2}}$, $T_{\frac{1}{2}}$ and $U_{\frac{1}{2}}$ vanish like $\delta M/M$, $\delta M^2/m_W^2$ and $\delta M^2/M^2$ as $\delta M \to 0$. For $M \gg m_W$, $T_{\frac{1}{2}}$ can be
the largest oblique parameter among the trio.
Thus for the physical quantities $\rho$ parameter and $W$ boson mass shift, we have
\begin{equation}
\Delta \rho_{\frac{1}{2}} \approx  
\frac{\hat \alpha}{12 \pi \hat s_W^2} N_C \frac{\delta M^2}{m_W^2} \; ,
\end{equation}
 and
\begin{flalign}\frac{ (\Delta m_W^2)_{\frac{1}{2}} }{m_Z^2}  
& \approx 
   {  \hat c^4_W \over  \hat c^2_W - \hat s^2_W }  \Delta \rho_{\frac{1}{2}}  \; ,
\end{flalign}
where the overall mass scale $M$ has  dropped out in both equations. 
Similar consequences can be obtained for the triplet case of $j=1$.
This explains the apparent hingum-tringum relationship between these two observables and the overall mass scale in all our plots.

\item 
There are always overlaps of the allowed orange regions from the $W$ boson mass shift and the allowed blue regions from the $\rho$ parameter (or $T$ parameter) using either the SM global fit adapted by the PDG 
or the new fit including the recent CDF II data.
Specifically, the allowed regions of mass splittings from $\Delta m_W \leq 75$ MeV overlap with those from $\rho$ parameter fit which gives $0.12<T<0.42$ when CDF II data was included. Meanwhile, the allowed regions of mass splittings from $\Delta m_W \leq 1$ MeV overlap with those from $\rho$ parameter fit which prefers $-0.09 <T< 0.15$ from the PDG. 
For the global fit that includes $\Delta m_W \leq 75$ (1) MeV from the CDF II data (PDG), the allowed mass splittings within all the electroweak multiplets we studied here are in the range of $15 \sim 50$ ($\sim$ a few) GeV.

\item 
Hypercharge $Y$ affects   the mass splittings very weakly as the plots are quite similar from representations with the same $N_c$  and $j$ but different $Y$ (compare color-singlets $(1,2)_{-3}$ and $(1,2)_{9}$ in Fig.~\ref{fig:doublet-colorsinglet}). The strongest constraint on the allowed regions of the mass splittings comes from the color factor $N_c$, as we can see from the figures that relatively high value of $N_c$ gives relatively small regions of allowed mass splitting parameters.  In the extreme cases 
of color-sextet/antisextet and color-octet, 
the mass splittings are confined to be less than a few GeV for $\Delta m_W \leq 1$ MeV from the SM global fit.

\item 
For scalar triplets, the allowed regions of the mass splitting $\delta M0$ between the mass of $\phi_{0}$ and $\phi_{-1}$ and the mass splitting $\delta M1$ between the mass of $\phi_{1}$ and $\phi_{-1}$ form rotated elliptic shapes on the $(\delta M0, \delta M1)$ plane. This is possibly due to the fact that we choose to measure the mass splittings from the mass of the ``lowest'' component $\phi_{-1}$.

\end{itemize}

\section{Conclusion}
\label{conclusion}

We have surveyed the general electroweak multiplets that can originate from a grand unified gauge theory such as $SU(5)$, $SO(10)$ or $E_6$. Besides the color (anti-)triplet leptoquarks, we can also have the color-sextet/antisextet and color-octet multiplets. Assuming these scalar electroweak doublets or triplets  survived at low energy, we computed the contributions from these multiplets to the electroweak oblique parameters and used them to constrain the mass splittings within each multiplet by the electroweak precision measurements of the $\rho$ parameter and $W$ boson mass. 

It is irresistible to extend our computation of the oblique parameters for general scalar electroweak multiplets to the vector case $V^\mu$. 
Vector LQs have been recently proposed to resolve the $B$-anomaly and muon $g-2$ discrepancy~\cite{Du:2021zkq}, as well as the CDF II $W$ boson mass anomaly and other phenomenological issues~\cite{Cheung:2022zsb}.
Since the only consistent theory for massive vector particles must be a Yang-Mills gauge theory with masses generated by spontaneously symmetry breaking via the Higgs mechanism, one expects the computation of oblique parameters, solely for a  vector multiplet,  will not be self-consistent without a full consideration of the gauge-Higgs sector. This implies a general analysis in parallel with the scalar case presented in this work is very unlikely, since here there was no need to consider the scalar potential that gives rise to their masses. Certainly one may proceed on a case-by-case basis where the vector multiplet is in definite irrep along with an appropriate Higgs sector. Results for the vector case will be reported elsewhere, as the complication involved in this case surpasses the scope of the present paper.

\begin{acknowledgments}
WYK thanks the hospitality by Kingman Cheung and Chong Sun Chu at National
Tsinghua University, Taiwan for the financial support and physics discussions on this work.
LC is supported in part by the National Key Research and Development Program of China under Grant No. 2020YFC2201501 and  the National Natural Science Foundation of China (NSFC) under Grant No. 12147103.
TCY is supported in part by the Ministry of Science and Technology of Taiwan under Grant No. 111-2112-M-001-035. 
\end{acknowledgments}

\section*{Appendix A -- Scalar Electroweak Multiplet Interaction Lagrangian}
\label{AppendixA}

Let the scalar multiplet denoted by a column vector $\phi$ with each of its components $\phi_m$ having mass $M_m$, where $m$ runs from $-j$ to $+j$ with $j$ being an integer or a half-integer, {\it i.e.},
$ \phi  = \left( \phi_j, \phi_{j-1}, \cdots, \phi_{-j+1},\phi_{-j} \right)^{\rm T}$.
The Lagrangian is 
\bea
\label{L-SVQ}
\mathcal L^S & = & \left( D_\mu  \phi \right)^\dagger \left( D^\mu  \phi \right) - \sum_{m=-j}^{+j} M_m^2 \phi^*_m \phi_m
\eea
with the covariant derivative $D_\mu$ is defined as
\be
\label{covD}
D_\mu = \partial_\mu - i g_s T^A G^A_\mu - i \frac{g}{\sqrt 2} \left( W^+_\mu t^+ + W^-_\mu t^- \right) - i \frac{g}{\cos \theta_W} Z_\mu \left( t^3 - \sin^2\theta_W Q \right) - i e Q A_\mu \; ,
\ee
where $T^A (A=1,\cdots,8)$, $t^a (a=1,2,3)$ are representations of the generators of $SU(3)_C$ and $SU(2)_L$ respectively for the scalar irrep, $t^\pm = \left( t^1 \pm i t^2 \right)$, $Q=t^3+Y$ is the electric charge (in unit of $e > 0$) with $Y$ the hypercharge, and $\theta_W$ is the Weinberg mixing angle. 

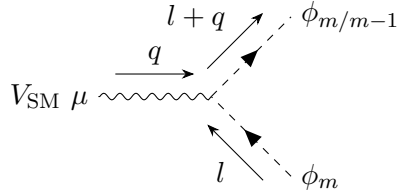
\begin{figure}[H]
        \centering
	\begin{tikzpicture}
\begin{feynman}
\vertex [label=left:\(V_{\rm SM}~\mu\)] (a);
\vertex [right=of a] (d);
\vertex [label=right:\(\phi_m\),below right=of d] (b);
\vertex [label=right:\(\phi_{m/m-1}\),above right=of d] (c);
\diagram* {
(a) -- [boson,momentum=$q$] (d),
(b) -- [charged scalar,momentum=$l$] (d),
(d) -- [charged scalar,momentum=$l+q$] (c),
};
\end{feynman}
\end{tikzpicture}
	\caption{\label{fig:AVV}
	Feynman rule for the cubic $V_{\rm SM} \phi \phi$ vertex is given by  $i C^m_{V_{\rm SM}} (2l+q)_\mu \delta_{\alpha\beta}$ where $V_{\rm SM}$ denotes $\gamma$, $Z$ or $W^-$, and $\delta_{\alpha\beta}$ is the 
    trivial color factor. The coupling coefficients are 
	$C^m_\gamma = e Q_m $, $C^m_{Z} = \frac{g}{\cos \theta_W} \left( m - \sin^2\theta_W Q_m \right) $, and $C^m_{W^-} = \frac{g}{\sqrt 2} \langle j \, m - 1 \vert t^- \vert j \, m \rangle = \frac{g}{\sqrt 2}\sqrt{(j+m)(j-m+1)}$ where $Q_m = m + Y$ with $Y$ being the hypercharge.}
\end{figure}

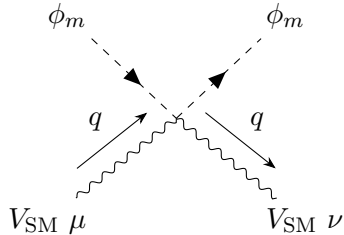
\begin{figure}[H]
        \centering
	\begin{tikzpicture}
\begin{feynman}
\vertex (b);
\vertex [above left=of b] (a) {\(\phi_m\)};
\vertex [above right=of b] (c)  {\(\phi_m\)};
\vertex [below left=of b] (d) {\(V_{\rm SM}~\mu\)};
\vertex [below right=of b] (e) {\(V_{\rm SM}~\nu\)};
\diagram* {
(a) -- [charged scalar] (b),
(b) -- [charged scalar] (c),
(d) -- [boson, momentum = $q$ ] (b),
(b) -- [boson, momentum = $q$ ] (e),
};
\end{feynman}
\end{tikzpicture}
	\caption{\label{fig:AAVV}
	Feynman rule for the quartic $V_{\rm SM}V_{\rm SM}\phi\phi$ vertex is given by $i C^m_{V_{\rm SM}V_{\rm SM}} g_{\mu\nu} \delta_{\alpha\beta}$ where $V_{\rm SM}$ denotes $\gamma$, $Z$ or $W^-$ and $\delta_{\alpha\beta}$ is the 
    trivial color factor. The coupling coefficients are $C^m_{\gamma\gamma} = e^2 Q_m^2 $, $C^m_{ZZ} = (\frac{g}{\cos\theta_W})^2 \left( m - \sin^2\theta_W  Q_m \right)^2 $, $C^m_{\gamma Z} = e \frac{g}{\cos\theta_W} Q_m \left( m - \sin^2\theta_W Q_m \right) $ and $C^m_{WW}=\frac{g^2}{2}\langle j \, m \vert \{t^+, t^-\} \vert j \, m \rangle = g^2 \left( j \left( j + 1 \right) - m^2 \right)$. }
\end{figure}

The relevant Feynman rules for computation of the oblique parameters for the scalar multiplet are depicted in Figs.~\ref{fig:AVV} and~\ref{fig:AAVV} for the cubic and quartic vertices respectively. Note that we have suppressed the color index in $\phi$ throughout the paper but explicitly included it in these Feynman rules.

\section*{Appendix B -- Useful Formulae}
\label{AppendixB}

The following integral
\begin{align}
B(M_1,M_2,q^2) & = \int_0^1 \left( x M_1^2 + (1-x)M_2^2 -x(1-x) q^2 \right) \nonumber \\
&\qquad  \times \log \frac{ x M_1^2 + (1-x)M_2^2 -x(1-x) q^2 - i 0^+  }{\mu^2} 
\end{align}
occurs frequently in all one loop vacuum polarization computation.
An analytic expression of this integral can be found in the literature, see for example~\cite{Bohm:1986rj}. 
However we only need the following two simple expressions for computation of the oblique parameters of scalar multiplet.
\begin{equation}
B\left(M_1,M_2,0\right)  = 
\frac{1}{2} \frac{M_1^4 \log M_1^2 - M_2^4 \log M_2^2}{M_1^2-M_2^2} - \frac{1}{4} \left( M_1^2 + M_2^2 \right) \left(1  + 2 \log \mu^2 \right) \; ,
\end{equation}
and 
\begin{align}
B^\prime \left(M_1,M_2,0\right) & = -\frac{1}{36} - \frac{1}{3} \frac{M_1^2 M_2^2}{\left( M_1^2 - M_2^2 \right)^2} - \frac{1}{12} \log \left( \frac{M_1^2}{\mu^2} \right) - \frac{1}{12} \log \left( \frac{M_2^2}{\mu^2} \right) \nonumber \\
& \qquad \qquad - \frac{1}{12} \frac{ M_1^6 - 3 M_1^4 M_2^2 - 3 M_1^2 M_2^4 + M_2^6 }{ \left( M_1^2 - M_2^2 \right)^3 } \log \frac{M_1^2}{M_2^2} \; ,
\end{align}
where $B^\prime(M_1,M_2,0)=d B(M_1,M_2,q^2)/d q^2\vert_{q^2=0}$.
In the event of $M_1=M_2=M$, we have
\begin{align}
B\left(M,M,0 \right) & = M^2 \log \frac{M^2}{\mu^2} \; ,\\
B^\prime \left(M,M,0 \right) & = - \frac{1}{6} \left( 1 + \log \frac{M^2}{\mu^2} \right) \; .
\end{align}


\end{document}